# Triplet Fusion Upconversion Nanocapsules for Volumetric 3D Printing

**Authors:** Samuel N. Sanders[a,b], Tracy H. Schloemer[a,b,c], Mahesh K. Gangishetty[b], Daniel Anderson[b], Michael Seitz[b], Arynn O. Gallegos[c], R. Christopher Stokes[b], and Daniel N. Congreve[b,c]*

**Affiliation:**

[a]These authors contributed equally.

[b]Rowland Institute at Harvard University, Cambridge, MA 02142

[c]Electrical Engineering, Stanford University, Stanford, CA 94305

*congreve@stanford.edu

**Abstract:** Two-photon photopolymerization delivers prints without support structures and minimizes layering artifacts in a broad range of materials. This volumetric printing approach scans a focused light source throughout the entire volume of a resin vat and takes advantage of the quadratic power dependence of two photon absorption to produce photopolymerization exclusively at the focal point. While this approach has advantages, the widespread adoption of two photon photopolymerization is hindered by the need for expensive ultrafast lasers and extremely slow print speeds. Here we present an analogous quadratic process, triplet-triplet-annihilation-driven 3D printing, that enables volumetric printing at a focal point driven by <4 milliwatt-power continuous wave excitation. To induce photopolymerization deep within a vat, the key advance is the nanoencapsulation of photon upconversion solution within a silica shell decorated with solubilizing polymer ligands. This scalable self-assembly approach allows for scatter-free nanocapsule dispersal in a variety of organic media without leaking the capsule contents. We further introduce an excitonic strategy to systematically control the upconversion threshold to support either monovoxel or parallelized printing schemes, printing at power densities multiple orders of magnitude lower than power densities required for two-photon-based 3D printing. The application of upconversion nanocapsules to volumetric 3D printing provides access to the benefits of volumetric printing without the current cost, power, and speed drawbacks. The materials demonstrated here open opportunities for other triplet fusion upconversion-controlled applications.

**Main text:** Three-dimensional (3D) printing, also known as additive manufacturing, has exploded in interest as new technologies have opened up a multitude of applications.[1–6] Stereolithography (SLA) is a particularly successful 3D printing approach where a photopolymer is patterned using light.[7] However, due to the linear absorption of the light, this technique requires photopolymerization to occur at the surface of the printing volume, necessitating interfacial printing. Creative work has found a number of ways to enhance this process, yet the interfacial nature of this printing process imparts fundamental limitations on resin choice and shape gamut. These constraints result in the use of extensive support structures[8,9] or the significant engineering of the printer system.[10,11]

One promising way to circumvent this interfacial paradigm is to move beyond linear absorption (Fig. 1A). Many groups have focused on utilizing two photon absorption (2PA) to print in a truly volumetric fashion.[3,7,12,13] The quadratic dependence of 2PA on incident light intensity allows for the generation of short wavelength photopolymerizing light only at the focal point of a laser beam, which can then be scanned in three dimensions to generate a 3D print (Fig. 1D). Utilizing 2PA, many groups and companies have been able to create remarkable nanoscale structures.[5,14] But the laser power (and thus cost and complexity) required to drive this process has limited print size and speed, preventing widespread application beyond the nanoscale. For instance, curing ~1 milliliter of resin using 2PA-based polymerization typically requires timescales on the order of tens of hours to days.[15,16] Additionally, the use of pulsed lasers with defined repetition rates introduces further complications, such as dielectric breakdown and resin boiling at intense light fields.[13] Thus, the next generation of volumetric 3D printing requires a quadratic process that supports continuous wave excitation, allows inexpensive implementation, and has low enough power requirements (<1 mW per voxel) to enable parallelization to thousands of voxels per exposure.

One promising alternative process to 2PA is triplet fusion upconversion.[17–20] This process takes advantage of excitonic states in annihilator molecules to generate an anti-Stokes emission relative to the sensitizer's absorption; see Fig. 1C for a full description of the process. Crucially, the final upconversion step requires collisions of two excited annihilator triplets, which fuse to form one higher energy annihilator singlet which then emits blue light that can be used to locally drive photopolymerization by coupling with a photoinitiator. This process has a quadratic nature due to the requirement for two triplets to meet, yet relatively low light fluences are required due to the high extinction coefficient of the sensitizer as compared to 2PA. Triplet fusion upconversion is also easily tunable in both excitation and emission wavelength with judicious selection of the sensitizer and annihilator.[17]

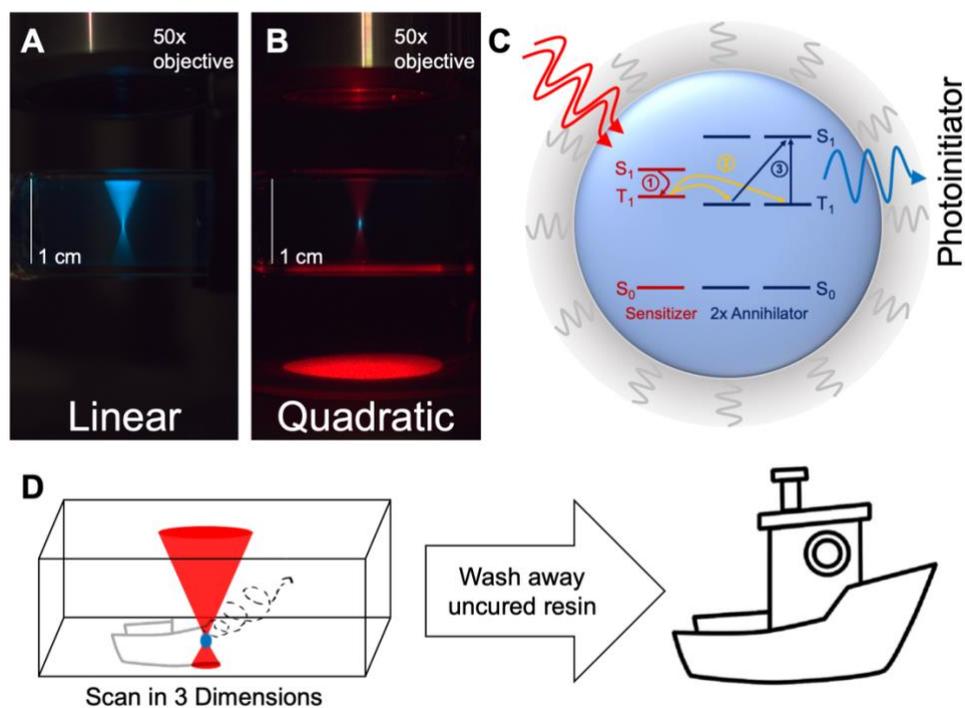

**Figure 1.** A-B) Comparison between linear and quadratic absorption processes in a 1 cm cuvette of PdTPTBP and Br-TIPS-anthracene in oleic acid. The linear process results from directly exciting the annihilator at 365 nm, and blue light is generated throughout the cuvette and visibly attenuates as a function of depth. The quadratic process results from triplet fusion upconversion by exciting the sensitizer at 637 nm, and blue light is generated only at the focal spot. The excitation sources are focused through a 50x objective. C) The upconversion process. Two low energy photons generate two singlet excitons on sensitizer molecules, which intersystem cross (1) to generate triplet excitons. These excitons then triplet energy transfer to the annihilator (2) where they undergo triplet fusion (3) to generate a higher energy singlet, which can radiatively decay by emitting a high energy photon that couples to the photoinitiator. (D) Cartoon depiction of the upconversion nanocapsule-facilitated printing process using monovoxel excitation. More details on the printing schematics are presented in the Supporting Information and Supporting Videos.

Triplet fusion upconversion is positioned to be an attractive analog to 2PA printing, see Fig. 1D and Fig. S2 for illustrated depictions of the printing schemes. However, when we initially attempted to use this process to drive 3D printing, we encountered several challenges. First, upconversion exhibits a threshold behavior: the upconversion efficiency crosses over from a quadratic to a linear dependence on input light above a certain fluence. Achieving control over this threshold value is crucial to applying upconversion to different printing schemes. For example, for a single-voxel printing approach, we targeted a focal point with a power density on the order of >100 W cm$^{-2}$. While this operating power is enormously smaller than the 10$^{12}$ W cm$^{-2}$ required for 2PA, it is considerably higher than the threshold value for typical upconversion systems.[17,21,22] On the other hand, using upconversion materials with a relatively low threshold would allow for low fluence, rapid, parallelized printing. Consequently, we needed systematic control over the threshold behavior of the upconversion materials.

Second, because triplet fusion upconversion relies on high concentrations of strongly absorbing molecules undergoing frequent collisions, direct addition of the sensitizer and annihilator to the resin posed severe practical restrictions. High concentrations of the molecules would need to be dissolved in the 3D printing resin, resulting in excessive attenuation of the input light and limited print volumes[23,24], analogous to the challenges observed in linear processes (Fig. 1A) where the blue light becomes less intense as it transits through the vial. Further, as the resin viscosity increases during printing, the rate of molecular collisions between the sensitizer and annihilator decreases. This will reduce the upconversion efficiency and result in a loss of the print selectivity. Taken together, we needed to engineer a sensitizer and annihilator pair such that we could access a wide range of upconversion threshold values, while simultaneously deploying them in a way to ensure both a high local concentration to maximize upconversion efficiency and a low global concentration of the upconversion materials to maximize light penetration depths.

To attack these challenges, we first developed a strategy to tune the UC threshold. We selected TIPS-anthracene (triisopropylsilyl ethynyl anthracene) as the annihilator and PdTPTBP (palladium (II) meso-tetraphenyl tetrabenzoporphine) as the sensitizer (Fig. 2A). This red-to-blue upconversion system works with attractive reported efficiencies of up to 30%.[21] However, the TIPS-anthracene threshold of 1.7 W cm$^{-2}$ was far too low to power our single-voxel printer with sufficient upconverted light and needed to be increased.

The threshold value of a triplet fusion system can be related to the nonradiative triplet decay rate $k_A$ and the rate of triplet fusion $k_{TF}$:[25,26]

$$I_{Threshold} \propto \frac{k_A^2}{k_{TF}}$$

To increase the threshold, we focused on increasing the triplet recombination rate, $k_A$, by adding heavy atoms to the molecule and subsequently increasing the threshold value, $I_{Threshold}$.[27] By adopting well-established acetylation chemistry, we were able to create a series of acenes with heavy atom substitution in order to adjust the threshold, see Fig. 2A and Supporting Information for synthetic methods.[28] The introduction of halogens to the anthracene core introduces little difference in emission or absorbance, yet the UC thresholds vary tremendously with the simple changes in the annihilator, see Fig. 2B and Fig. S3. In Fig. 2C, we plot the relative upconversion efficiency of each of these acenes against power density. This relative efficiency is the derivative of the upconverted light versus the input light, and it increases at low powers until the threshold, when it plateaus and becomes constant. Indeed, this relationship between heavy atom substitution and measured UC thresholds is observed for all five molecules. Importantly, the measured thresholds range from 1.7 W cm$^{-2}$ for unsubstituted TIPS-anthracene all the way up to 283 W cm$^{-2}$ for the 2Br-TIPS-anthracene, spanning over two orders of magnitude. This molecular library imparts enormous flexibility in printer design: a monovoxel excitation source can take advantage of the high threshold of Br-TIPS- or 2Br-TIPS-anthracene annihilators (Fig. 1D), while a large area parallel excitation printer, which requires lower excitation intensities to drive thousands of voxels simultaneously, could use the Cl-TIPS-anthracene or even the unsubstituted TIPS-anthracene annihilators (Fig. S2).

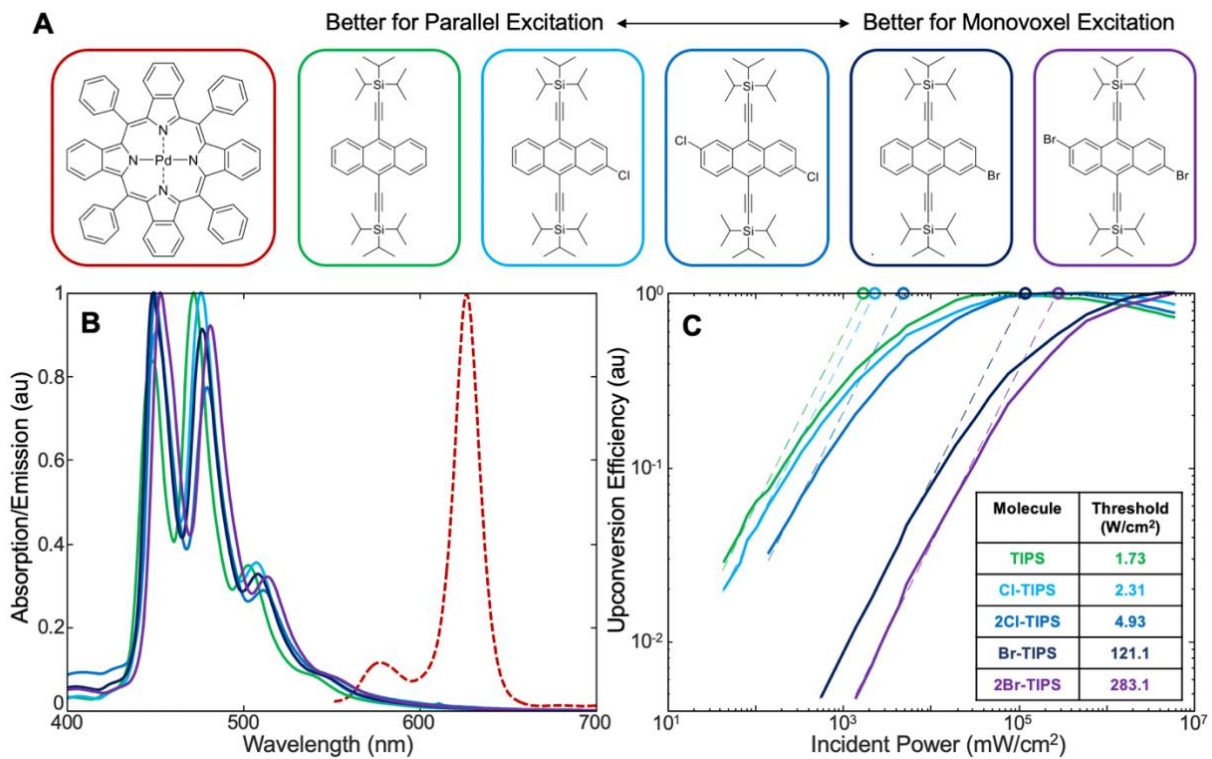

**Figure 2.** A) From left to right, the chemical structures of PdTPTBP, TIPS-anthracene, Cl-TIPS-anthracene, 2Cl-TIPS-anthracene, Br-TIPS-anthracene, and 2Br-TIPS-anthracene. B) The absorption of PdTPTBP (dashed red) and the photoluminescence of the annihilators. C) The upconversion efficiency as a function of input power. A linear fit (dotted lines) to the quadratic

regime gives the threshold of each material (circles). See the Supporting Information for sample preparation details.

With the upconversion materials now in hand, we turned to the challenge of their deployment. To achieve this, we sought to encapsulate nanodroplets of the upconversion stock solutions. We targeted nanoscale dimensions to minimize the optical scatter of the printing resin. While we and others have previously built upconverting micelles out of block copolymers for applications in aqueous solutions, we found that these micelles were unstable in organic solvents and rapidly released their contents, resulting in a loss of upconversion due to inefficient molecular collisions at low concentrations.[29–32] Instead, we sought a nanoencapsulation that would disperse in organic-based 3D printing resins without leaking the contents or scattering light. We were inspired by recent work from Kwon et al,[29] who built substantially more durable upconverting silica nanocapsules relative to prior reports.[29–32] Even so, we found that these materials significantly scattered the input laser beam due to aggregation of the nanocapsules during shell formation, and were not dispersible in resins (Fig. S4). To overcome these challenges, we designed a nanocapsule synthesis that incorporated a long poly(ethylene glycol) (PEG) chain as a solubilizing ligand on the exterior of the silica shell. However, we found that we could not rely on the physical adsorption of PEG chain to be a durable stabilizing ligand. Incorporating a silane-terminated PEG, which can covalently graft to the silica shell of the nanocapsule,[33] was crucial to prevent aggregation over time and to allow the nanocapsules to disperse without scatter in 3D printing resins (Fig. S4). Full details of the optimized nanocapsule synthesis can be found in Fig. 3A and the Supporting Information. Collectively, this new nanoencapsulation system is a significant advance to prior reports, as it is inherently modular with regard to the capsule contents and solution-state compatibility (*vide infra*).

Electron microscopy of the resulting upconverting nanocapsules (UCNCs) (Fig. 3C and 3D) shows uniform capsules approximately 50 nm in diameter. When dispersed in water, the effective diameter is approximately 75 nm as determined by dynamic light scattering (Table S1). Of particular importance is the stability imparted by the silica shell; to demonstrate this, we diluted the nanocapsule solution from both the initial micelle formation after dropwise addition of (3-aminopropyl)triethoxysilane (APTES) and the final shelled UCNCs at 100:1 in acetone. In the former case, the micelles fall apart due to inefficient triplet energy transfer from the sensitizer to annihilator upon dissolution of the micelle, resulting in limited upconversion and significant PdTPTBP phosphorescence at ~800 nm. In the latter case, the bright upconversion from the UCNCs is preserved with minimal sensitizer phosphorescence, Fig. 3E. In addition to the durability improvements, we highlight the negligible phosphorescence observed in the UCNCs, which is extremely challenging to suppress in encapsulated systems[31] and signifies efficient collisions between sensitizer and annihilator. We were further able to disperse the capsules in a number of common solvents while maintaining bright upconversion, see Fig. S5. UCNCs dispersed in acrylic acid and stored in the dark for >24 hours suggest good durability over time, as the upconversion photoluminescence remains constant as compared to a freshly prepared sample. The thermal stability is also an important consideration due to the exothermic nature of chain growth reactions in the 3D printing process. Thermogravimetric analysis of the UCNCs (Fig. S6) shows that the decomposition onset temperatures of the UCNCs and each individual constituent are well above the expected printing temperatures (> 200 ºC for the sensitizer, annihilator, and silane-terminated PEG). Additionally, we found approximately half of the weight of the purified nanocapsule paste is composed of water, which later influenced the resin design (*vide infra*).

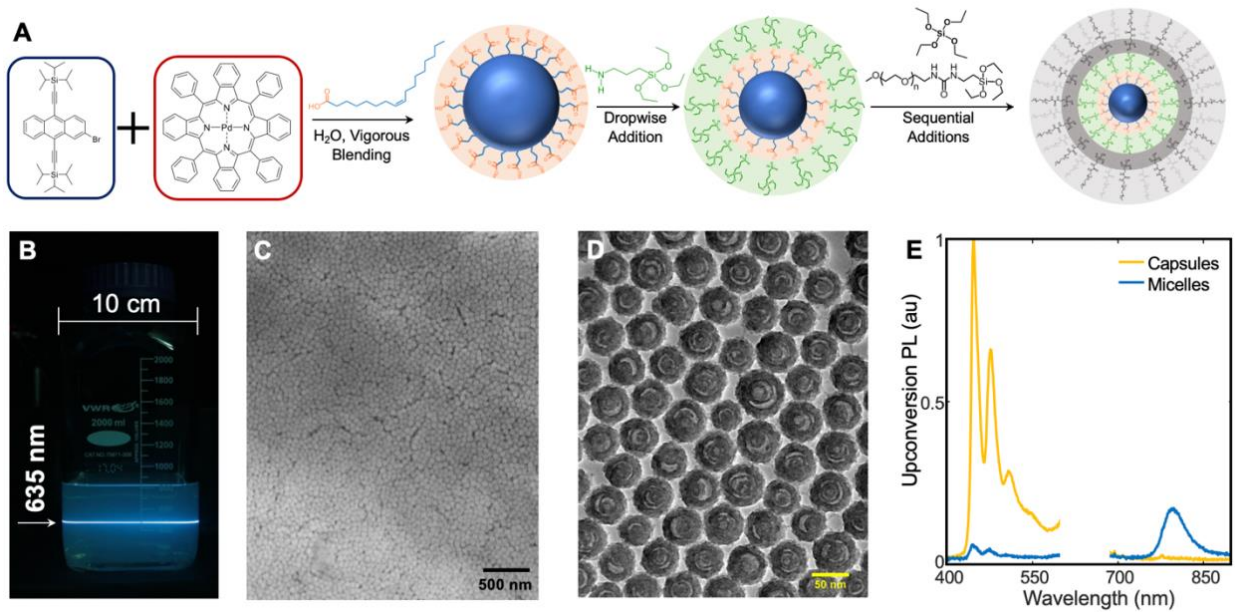

**Figure 3.** A) Overview of the UCNC synthesis with Br-TIPS-anthracene as the annihilator. B) UCNCs diluted in acetone show long light penetration depths. The capsules are excited at 635 nm and are imaged through a 600 nm shortpass filter. C) SEM of the UCNCs shows the scale and uniformity of the nanoparticle synthesis. D) TEM of the UCNCs. E) Dispersion of the initial micelles (blue) and final UCNCs (orange) in acetone under the same conditions shows the necessity of the silica shell to upconversion survival. The emission peak at ~800 nm corresponds to phosphorescence from the sensitizer PdTPTBP. See the Supporting Information for sample preparation details.

We then introduced our capsules to a simple 3D printing resin consisting of all commercially available materials; the full fabrication details are presented in the Supporting Information. First, we utilized acrylic acid (AA) to disperse the nanocapsules in an acrylate-based monomer. A polar protic monomer was required to disperse the UCNCs due to the moderate water content of the capsule paste and polar PEG chains affixed to the nanocapsule surface. We then used pentaerythritol tetraacrylate (PETA) to introduce the capsules into a highly cross-linkable matrix. The final optimized resin contains ~15 wt.% upconversion nanocapsules. We affirm that the high UC nanocapsule loading does not appreciably impact the ability to print deep within a vat, as the optimized resin transmits >85% of 637 nm light through a 1 cm polystyrene cuvette.

Control of the free radical polymerization and voxel volume was accomplished with a combination of both chemical and optical control. With regard to chemical control, we incorporated four resin additives: the photoinitiator, radical inhibitor, light blocker, and viscosity tuner. We used the commercially available photoinitiator Ivocerin (bis-(4-methoxybenzoyl)diethylgermanium) to absorb the blue UC emission and initiate polymerization (Fig. S7). We find that the addition of a radical inhibitor, TEMPO (2,2,6,6-Tetramethylpiperidin-1-yl)oxyl), at 3 ppm improves the print resolution by preventing significant polymerization outside of the irradiated voxel by efficiently terminating carbon-carbon radicals.[7,34] During the TEMPO concentration optimization, we found that at constant printing power, the TEMPO concentration is directly proportional to the print speed required for comparable prints. We also added a broad-spectrum light blocker, Sudan I,[35] to sharpen the resolution by attenuating the upconverted light before it travels substantially far from the focal point (Fig. S7). Ultimately, we find that the concentrations of Ivocerin, TEMPO, and Sudan 1 can be fine-tuned depending upon the desired print power, speed, and resolution. Finally, the resin viscosity is further increased with the addition of a hydrophilic fused silica to generate self-supporting prints (Aerosil 200, see Video S1).

For monovoxel printing, by overfilling the back aperture of a 0.55 NA objective with a collimated laser beam and mounting the objective onto a fused deposition modeling printer, we were able to trace a well-defined focal point in space throughout resins using Br-TIPS-anthracene as the nanoencapsulated annhilator; Fig. 1D, Fig. S1, Fig. S8, and Video S2. Due to the quadratic power dependence of triplet fusion, the upconversion is active predominantly at the focal point of the laser where the light is most intense, resulting in a well-confined blue voxel that drives local photopolymerization. Control resins did not print within the power range and time required for printing (resins without upconversion nanocapsules, or resins with upconversion nanocapsules but without a photoinitiator).

A standard yet difficult test of 3D printing systems is the benchmark boat print (often referred to as Benchy[36]), shown in Fig. 4. In a proof-of-concept for printing with this technology, we faithfully reproduce this print at small scales using the monovoxel printing setup and the optimized Br-TIPS-anthracene-based resin. Printing without support structures simplifies post-processing and limits surface blemishes: see Fig. S9 to compare with the same file printed on a commercial printer where >80% of the resin is used in the support structure. The power and print speed were carefully optimized to prevent "underprinting" and "overprinting" (Fig. S10) to realize high fidelity, reproducible prints in under two hours (Fig. 4 and Fig. S11). We print at power densities of ~$4.5 \times 10^3$-$4.9 \times 10^3$ W cm$^{-2}$ (Video S2 and Fig. S8), confining linear upconversion to within 80 μm from the voxel's focal point (Supplemental Note 1). We further exemplify the fine detail attainable by printing the highly intricate Harvard University logo (Fig. S12, Video S3).

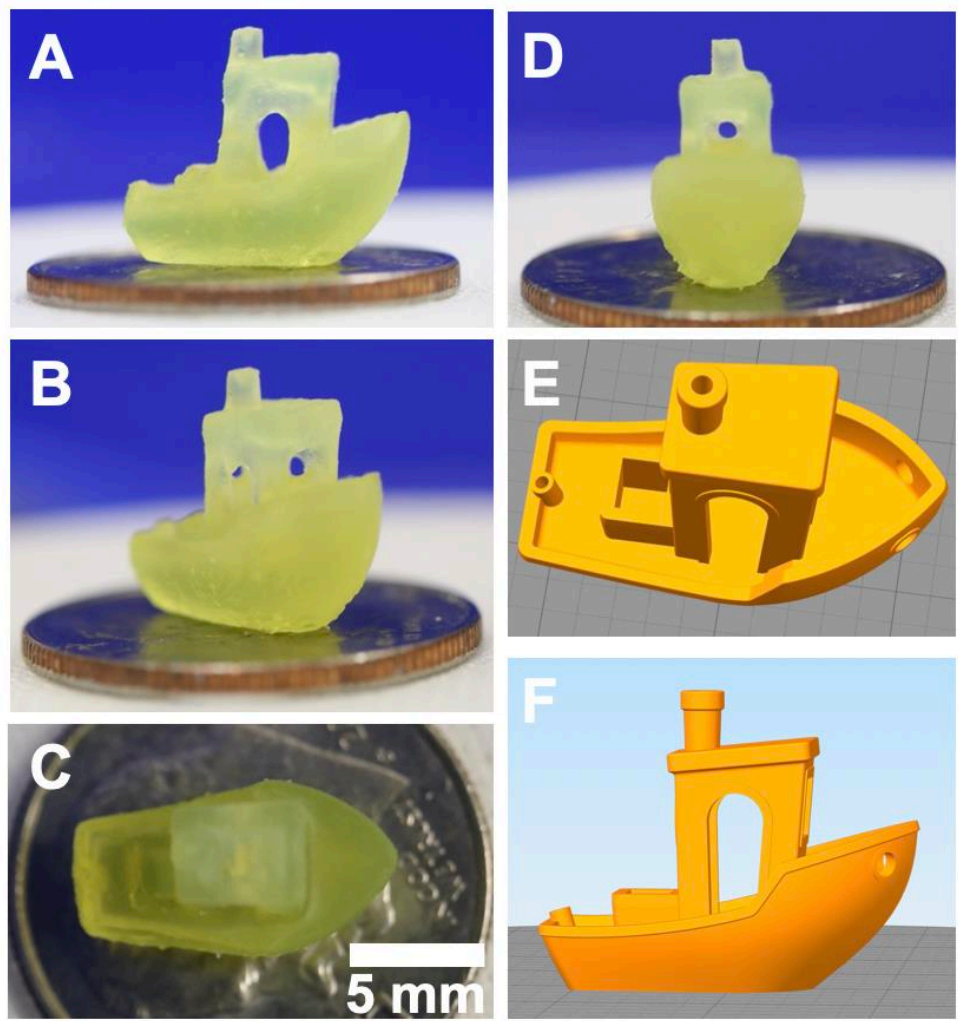

**Figure 4.** A-D) Side and top views of our final benchmark boat (Benchy) print, sitting on a dime for scale. E-F) A side and top view of the Benchy STL (Standard Tessellation Language) file. The side and top views of the final print show the faithful reproduction of the main features. See the Supporting Information for sample preparation details.

We also demonstrate large area parallel excitation printing to generate an intricate gear using the low UC threshold TIPS-anthracene-based resin (Fig. S2, Fig. S8, Video S4). We print a detailed gear and a Stanford University logo using an LED patterned by a digital micromirror device (DMD) printer in 8 minutes at 78 mW cm$^{-2}$ (Fig. 5, Fig. S13). We present the full print and washing process in Video S4. Microscope images show minimal surface blemishes on the gear, even after the excess resin was washed away (Fig. S14). We could not generate a print with an upconversion nanocapsule-free control resin, nor with a BrTIPS-anthracene-based resin due to the high upconversion threshold, further highlighting the importance of matching the appropriate threshold UC system to a specific printing motif.

We then sought to quantify the smallest features attainable with our optimized resin using the DMD projection system and a mask to pattern lines, see the Supporting Information for details. We reproducibly printed features as small as 50 μm at a power density of 224 mW cm$^{-2}$ at the highest energy density achievable with our set up (Fig. 5, Table S2). We anticipate that smaller features could be generated with a combination of improved optics and resin formulation (e.g., increased overlap between the upconversion emission and photoinitiator absorption).

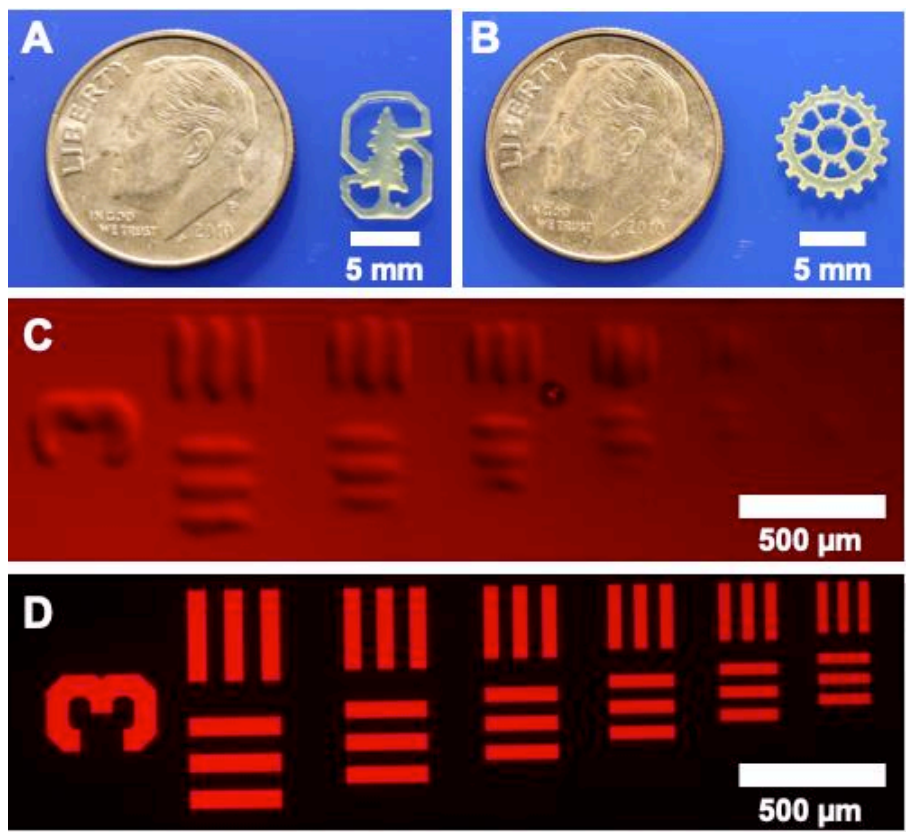

**Figure 5.** Top views of Stanford logo (A) and gear (B) prints, sitting next to a dime for scale. Microscope images of lines printed in ~0.1 mL of resin (C) with light projected through the United States Air Force Target Test mask (D). See the Supporting Information for sample preparation details.

The threshold tuning across a wide range of power densities and the encapsulation for good dispersibility are critical milestones for upconversion-facilitated lithography. While there are reports of photopolymerization using upconverted light (Table S3),[18,37–46] the shape gamut versatility and range of power densities used are unique to this work. Challenges observed in other systems are likely due to a combination of factors depending upon the upconversion mechanism (e.g., the high power densities required for UC with inorganic nanoparticle, the high molar absorptivity of organometallic sensitizers for triplet-fusion UC). Moreover, inorganic nanoparticle synthesis can be challenging to scale, which may further complicate the shape gamut versatility even with improved optics and resin formulations. Overall, no other reported UC-facilitated photopolymerization demonstrates the scope and versatility as this UC NC system provides.

Using a process with quadratic dependence on light intensity and low-threshold nanocapsules, we demonstrate the ability to arbitrarily pattern light and cure considerable volumes of resin as compared to other quadratic processes. There is critical need for advancing the processing speed of volumetric printing to deliver intricate resolution prints in practical timescales.[15] While the optical and algorithmic engineering required to achieve 3D selectivity with multi-voxel printing is beyond the scope of the current work, we demonstrate curing of considerable volumes of resin in minutes instead of tens of hours required using 2PA-based printing.[15,16,47] Further, this approach can print without support structures, with lower energy laser inputs, with greater materials selection versatility, and with higher speeds as compared to 2PA.

This UCNC printing platform suggests promise in printing channels, fine filigree features, and other shapes that may be challenging for traditional macroscale SLA approaches. This technique enables printing in resins that cannot be printed in a traditional SLA 3D printing setup, such as resins with high viscosity, resins requiring air-free polymerizations techniques[48], or prints with soft or flexible parts. Moving forward, optical engineering of projection-based systems to achieve high numerical aperture over large field of view will allow access to increasingly complex 3D prints at 100,000s of pixels at a time using continuous wave, low energy lasers: 3D printing without steps between layers.[15,47] We expect the ability to tune the threshold behavior over orders of magnitude to facilitate tailoring of these UCNCs to a variety of printing excitation schemes beyond the simple monovoxel or parallel excitation printers demonstrated here, such as full projector-based printing approaches, which can enable a variety of applications that require rapid, customizable, precision 3D printing. Finally, we expect to combine this technique with recent technological developments in optical parallelization to greatly increase print speeds.[49,50] Other upconversion systems can likely yield successful prints in a similar fashion to enable new lithographic processes, with a different set of application-based tradeoffs for their implementation (e.g., excitation wavelength, irradiation dosage, resin formulation).[41]

This demonstration shows the strength of integrating triplet-fusion upconversion nanocapsules for volumetric 3D printing towards fourth generation additive manufacturing. The ability to simply exchange the UC nanocapsule contents is a critical feature that enables discrete, localized photochemistry for many applications within the realm of 3D printing. We believe UC nanocapsules can serve as a key enabling technology for a variety of optically-controlled systems that currently suffer from high energy light penetration limitations.

**Acknowledgements:** This research is funded through the support of the Rowland Fellowship at the Rowland Institute at Harvard University, the Harvard PSE Accelerator Fund, and the Gordon and Betty Moore Foundation. A portion of this work was performed at the Harvard Center for Nanoscale Systems (CNS), a member of the National Nanotechnology Coordinated Infrastructure Network (NNCI), which is supported by the National Science Foundation under NSF, Award No. 1541959. A portion of this work was performed at the Stanford Nano Shared Facilities (SNSF), supported by the National Science Foundation under award ECCS-2026822. A portion of this work was performed at the Stanford ChEM-H Macromolecular Structure Knowledge Center.

SNS acknowledges the support of the Arnold O. Beckman Postdoctoral Fellowship. MS acknowledges financial support from the Swiss National Science Foundation (Project No. P1SKP2 187676). We thank Professor C. Jeffrey Brinker from the University of New Mexico and Dr. Swaroop Kommera from Stanford University for fruitful discussions. We thank Dr. Victor A. Lifton from Evonik for supplying the Aerosil 200, and Dr. Alan Sellinger and Dr. Allison Lim from Colorado School of Mines for performing the TGA experiments.

The STL file for 3DBenchy - The jolly 3D printing torture-test by CreativeTools.se by CreativeTools is licensed under the Creative Commons - Attribution - No Derivatives license. The gear we designed and printed was inspired by Gear 1 from the Wikimedia Commons, and is licensed under the Creative Commons – Attribution-Share Alike 3.0 Unported license.

**Competing Financial Interests:** Harvard University has filed several patents based on this work. SNS, RCS, and DNC are co-founders of Quadratic3D, Inc. SNS is the Chief Technological Officer, DNC is the Chief Scientific Advisor, and RCS is an advisor to Quadratic3D, Inc.

**Supporting Information:** Materials and Methods, Fig. S1 – S14, Tables S1-S3, Supplemental Note 1, Supplemental Files S1-S2, Supplemental Videos S1-S4.

**Data availability:** The data supporting this study are available from the corresponding author upon reasonable request.

# Supporting Information

**Triplet Fusion Upconversion Nanocapsules for Volumetric 3D Printing**

**Authors:** Samuel N. Sanders[a,b], Tracy H. Schloemer[a,b,c], Mahesh K. Gangishetty[b], Daniel Anderson[b], Michael Seitz[b], Arynn O. Gallegos[c], R. Christopher Stokes[b], and Daniel N. Congreve[b,c]*

**Affiliation:**

[a]These authors contributed equally.

[b]Rowland Institute at Harvard University, Cambridge, MA 02142

[c]Electrical Engineering, Stanford University, Stanford, CA 94305

*congreve@stanford.edu

**Materials and Methods**

**Materials:** All chemicals were used as received. (Triisopropylsilyl)acetylene, n-butyllithium (2.5M in hexanes), (3-aminopropyl)triethyoxysilane (APTES), pentaerythritol tetraacrylate (PETA), 1-phenylazo-2-naphthol (Sudan 1), 2,2,6,6-Tetramethyl-1-piperidinyloxy (TEMPO), poly(ethylene glycol) diacrylate (PEGDA) and acrylic acid were purchased from Sigma Aldrich. PdTPTBP was purchased from Frontier Scientific. 99% oleic acid was purchased from Beantown Chemical. Anhydrous tetraethyl orthosilicate (TEOS) was purchased from Acros Organics. 10K MPEG-Silane was purchased from Nanosoft Polymers. Bis(4-methoxybenzoyl)diethylgermanium (Ivocerin) was purchased from Synthon Chemicals GmbH & Co. 2,6-dichloroanthracenequinone was purchased from HAARES ChemTech. Aerosil 200 was provided by Evonik Industries.

**Annihilator Synthesis:** TIPS-anthracene (9,10-bis((triisopropylsilyl)ethynyl)anthracene) was purchased commercially from Sigma Aldrich, and bromo-TIPS-anthracene (2-bromo-9,10-bis[((triisopropyl(silyl)ethynyl]anthracene) and dibromo-TIPS-anthracene (2,6-dibromo- 9,10-bis[((triisopropyl(silyl)ethynyl]anthracene) synthesis have been reported previously.[1–3] The chloro-TIPS-anthracene and dichloro-TIPS-anthracene were synthesized according to the same protocol.

**Chloro-TIPS-anthracene** (2-chloro-9,10-bis[((triisopropyl(silyl)ethynyl]anthracene)

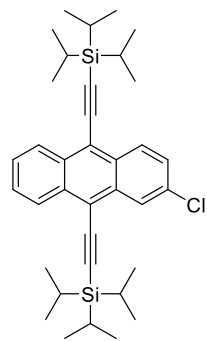

To a solution of 8.21 mL (36.6 mmol, 3.5 eq) of TIPS-acetylene in 10 mL of dry tetrahydrofuran in inert atmosphere was added at 0 °C 13.8 mL of 2.5M *n*-butyllithium in hexanes (34.5 mmol, 3.3 eq). This solution was kept at 0 °C for 30 minutes, then 2.54 grams (10.4 mmol, 1 eq) of 2-chloro, 9,10-anthracenequinone was added. The solution was allowed to warm to room temperature over the course of an hour when 6.00 grams of $SnCl_2*2H_2O$ (26.6 mmol, 2.6 eq) was added along with 1 mL of 10% aqueous hydrochloric acid and the reaction was stirred another hour. The resulting solution was extracted between hexanes and water three times, then solvent and residual TIPS-acetylene were removed *in vacuo*. Column chromatography on silica gel (hexanes and dichloromethane) yielded products as bright yellow solids.

1.83 grams (31% yield)

$^1$H NMR (600 MHz, CDCl$_3$, δ ppm): 8.66 (d, 1H), 8.62 (m, 2H), 8.66-8.63 (m, 2H), 8.59 (d, 1H), 7.63 (m, 2H), 7.55 (m, 1H), 1.28 (m, 42H)

$^{13}$C NMR (125 MHz, CDCl$_3$, δ ppm): 135.80, 135.47, 135.45, 135.07, 133.25, 131.72, 130.62, 130.04, 129.97, 129.90, 129.76, 128.60, 121.65, 120.55, 108.24, 108.11, 105.50, 105.43, 21.53, 14.17, 14.15

MS (ESI): Calculated m/z: 572.305; Observed m/z: 572.3066

**Dichloro-TIPS-anthracene** (2,6-dichloro -9,10-bis[((triisopropyl(silyl)ethynyl]anthracene)

To a solution of 8.21 mL (36.6 mmol, 3.5 eq) of TIPS-Acetylene in 10 mL of dry tetrahydrofuran in inert atmosphere was added at 0 °C 13.8 mL of 2.5M *n*-butyllithium in hexanes (34.5 mmol, 3.3 eq). This solution was kept at 0 °C for 30 minutes, then 2.90 grams (10.4 mmol) of 2,6-dichloro, 9,10- anthracenequinone was added. The solution was allowed to warm to room temperature over the course of an hour when 6.00 grams of SnCl$_2$*2H$_2$O (26.6 mmol, 2.6 eq) was added along with 1 mL of 10% aqueous hydrochloric acid and the reaction was stirred another hour. The resulting solution was extracted between hexanes and water three times, then solvent and residual TIPS-acetylene were removed *in vacuo*. Column chromatography on silica gel (hexanes and dichloromethane) yielded products as bright yellow solids.

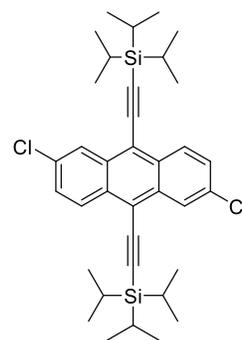

2.14 grams (34% yield)

$^1$H NMR (600 MHz, CDCl$_3$, δ ppm): 8.62 (d, 2H), 8.54 (d, 2H), 7.55 (m, 2H) 1.28 (m, 42H)

$^{13}$C NMR (125 MHz, CDCl$_3$, δ ppm): 136.13, 135.44, 133.60, 131.66, 131.20, 128.67, 120.84, 108.87, 104.92, 21.49, 14.13

MS (ESI): Calculated m/z: 606.2666; Observed m/z: 606.2676

**Upconversion Nanocapsule Synthesis:** Fresh upconversion stock solutions were prepared for each batch of capsules. Under red lighting, saturated solutions of the sensitizer (PdTPTBP, 2 mg mL$^{-1}$) and annihilator (TIPS-anthracene, 8 mg mL$^{-1}$, or Br-TIPS-anthracene, 70 mg mL$^{-1}$) were prepared in 99% oleic acid at room temperature, and the mixtures were allowed to stir for at least 4 hours before filtering with a 0.45 μm PTFE filter. Red lighting was used to prevent demetallation of PdTPTBP.[4] Stock solutions for TIPS-anthracene capsules were 2:1:2 TIPS-anthracene:PdTPTBP:OA by volume, and stock solutions for Br-TIPS-anthracene capsules were 0.5:1:3.5 Br-TIPS-anthracene:PdTPTBP:OA by volume. After the stock solutions were prepared, rigorous red lighting was no longer required and ambient lighting was used. Milli-Q water (200 mL) was chilled over an ice bath for at least one hour (temperature ~5 ºC) and then poured into to a Vitamix Blender (Amazon.com) in a nitrogen filled glovebox. The upconversion stock solution containing sensitizer and annihilator (1.45 mL) was carefully dispensed into the water in one portion. The solution was blended for 60 s at the maximum speed, and the emulsion was transferred to a 500 mL 2 necked round bottom flask and immediately stirred at high speed with an egg-shaped stir bar (~1000 rpm). (3-aminopropyl)triethoxysilane (0.75 mL) was added until the mixture became transparent, and then 10K MPEG-Silane (4 g) was immediately added to prevent capsule aggregation and stirred vigorously to disperse evenly. Within 5 minutes of the MPEG-Silane addition, anhydrous tetraethyl orthosilicate (TEOS, 30 mL) was added in one portion. The flask was sealed and the solution stirred vigorously at room temperature. After approximately 10 mins, the flask was sealed before removing from the glovebox and was stirred vigorously at 65 ºC under constant nitrogen pressure connected to a Schlenk line. A second portion of MPEG-Silane (4 g) was added after ~42 hours. After ~48 hours, the reaction crude was cooled to room temperature under constant nitrogen pressure, poured into a centrifuge tube, and centrifuged at 8670 x g (Fisher Lynx Sorvall centrifuge) for one hour at room temperature (18-22 ºC), after which the pellet was discarded. The supernatant was further centrifuged at 8670 x g for 12-14 hours at room temperature. After the second centrifuge, the paste-like UCNCs (~8-10 g) were immediately transferred to the glovebox. See Fig. 3 and Fig. S6 for representative characterization (SEM, TEM, photoluminescence, TGA). For this work, we have scaled this synthesis by 2-3x using the Vitamix blender with no apparent change in nanocapsule quality or physical/optical properties.

**Measurements:**

Photographs were taken with a with a Canon EOS Rebel T6i except where noted. The time lapse videos were recorded with an iPhone 8 and/or iPhone 12. All images presented are unedited.

Fig. 1 images are of a dilute stock solution of PdTPTBP and Br-TIPS-anthracene. The quadratic voxel was generated with 637 nm light. The linear voxel was generated from a fiber coupled 365 nm LED coupled into the same optical path. The contents in the linear cuvette were diluted by a factor of 2 relative to the contents in the quadratic cuvette to better show the entire voxel. The spot size was measured by moving a razor blade through the spot with a micrometer.

Fig. 1, 3B, and S5 images are taken with the same camera with the sample excited in free space by a 635 nm laser from the right through a 550 nm short pass filter. F127 micelles were made as discussed previously[5] with the sole change that Br-TIPS-anthracene was used as the annihilator. Capsules and micelles were diluted 30:1 in the specified solvent. The tap water was taken directly from the sink and left uncapped for 20 minutes before the image was taken. All other cuvettes were mixed in the glovebox. The upconversion photoluminescence spectra presented in Fig. 3E has the laser scatter omitted (~635 nm).

Absorption spectra were collected on a Cary-5000 UV-Vis spectrometer.

Photoluminescence spectra were recorded with an Ocean Optics QE Pro. For intensity dependence presented in Fig. 2, incident laser intensity was measured with a calibrated Si photodetector from Newport and varied using ND filters. The emission intensity was measured with the QE Pro spectrometer and integrated. No variation in emission shape or time was observed throughout the measurement.

SEM image presented in Fig. 3 was captured by using an in-lens (immersion lens) detector on Supra55VP Field Emission Scanning Electron Microscope (FESEM) at 10 keV.

TEM image presented in Fig. 3 was captured by JEOL-2100 HR-TEM operated at 200 kV. The sample was drop casted on a polymer coated Cu grid.

Thermogravimetric analysis presented in Fig. S6 was performed using a TA instruments TGAQ500 with high resolution sensitivity at a ramp rate of 50 °C min$^{-1}$ with 4.00 °C resolution under nitrogen flow up to 1000 °C.

Dynamic light scattering presented in Table S1 was collected on a Brookhaven Instruments 90Plus Nanoparticle Size Analyzer. The temperature was held at 25 °C with a 5 s equilibration time and 3 total measurements. A sample of the supernatant was collected after the first centrifuge step during purification, diluted 10x in MilliQ water, and filtered with a 0.2 um PVDF syringe filter into a 1 cm polystyrene cuvette. Nanoparticle aggregation was not observed when analyzing samples with an order of magnitude difference in concentration.

A Thorlabs Power meter PM100D with a S120VC photodiode was used to measure the voxel power as a function of the laser dial setting as used on the FDM printer and the focal plane power as used on the DMD printer.

Images presented in Fig. 5C were taken in an inverted microscope (Nikon Ti-2 Eclipse) with a 4x magnification illuminated with a red lamp. Printing images on the microscope were taken with a Kiralux™ 8.9 Megapixel Color CMOS Camera purchased from ThorLabs.

Images presented in Fig. S14 was taken in an inverted microscope (Nikon Ti-2 Eclipse) with a 4x magnification illuminated under white light.

**Monovoxel Excitation Using the FDM Printer:**

<u>Br-TIPS-anthracene Resin Preparation (10 mL resin):</u> All resins were prepared in a nitrogen atmosphere under red light in a 20 mL scintillation vial covered in aluminum foil. To 150 mg of Aerosil 200 silica (A200), 5 mL pentaerythritol tetraacrylate (PETA) was added and the mixture was stirred at 90 ºC for 15 minutes, and then the temperature was reduced to 60 ºC and the solution was allowed to cool for 30 minutes. To the hot PETA/A200 solution, TEMPO (30 μL of a freshly prepared 1 mg mL$^{-1}$ acrylic acid solution diluted from 10 mg mL$^{-1}$) and Sudan 1 (40 μL of a freshly prepared 1 mg mL$^{-1}$ acrylic acid solution diluted from 10 mg mL$^{-1}$) were added. The solution was inverted to mix before adding Ivocerin (550 mg) and 500 μL acrylic acid. The solution was stirred at 60 ºC for 30-45 minutes, to ensure the solution is optically transparent. Then, the capsule paste solution (2.5 mL of a 0.67 g mL$^{-1}$ in acrylic acid, stirred for >1 hour at room temperature) and acrylic acid (1 mL) were added. After shaking the resin, PETA was added for a total resin volume of 10 mL. The solution was then sparged under nitrogen for 15 minutes at 60 ºC. The resin was allowed to stir and cool to room temperature on the warm hot plate for one hour. The resin was then poured into a polystyrene cuvette (3.5 mL, 1 inch pathlength, sealed with parafilm), mixed in a FlackTek SpeedMixer to remove bubbles (1000 rpm, 5-8 minutes), and used immediately for printing.

<u>Printing</u>: The printing is performed on a custom built, highly modified Kossel Delta configuration reprap printer, with many modifications gathered from Haydn Huntley (https://www.kosselplus.com/) (Fig. S1). The firmware run on the printer is a fork of Reprap: RepRapFirmware-dc42, available at https://github.com/dc42/RepRapFirmware, and the printing electronics are the Duet 2 Wifi controller (https://www.duet3d.com/DuetWifi).

STL files are sliced in Simplify3D to generate gCode inputs to the printer. The Benchy gCode file is provided as a Supplemental File.

The printing is powered by a Thorlabs S4FC637 637 nm 70 mW fiber coupled laser (Fig. S8). The laser is collimated with a 20 mm focal length lens then fed into a 50X Mitutoyo Plan Apochromat Objective where it is focused into the resin. The entire optical system is moved by the printer in three dimensions to generate the print. Benchy is printed from the bottom of the cuvette to the top (Video S2). The print starts at a laser power setting of 16-17 mW. The laser power setting is reduced to 15-16 mW after one hour. This corresponds to an output power of <4 mW at the focal point; we measure power to be between 3.6 and 3.9 mW, and power densities of $4.5 \times 10^3$-$4.9 \times 10^3$ W cm$^{-2}$ based on a 10.1 μm measured diameter of the focal point. The print is run at a speed of 50% relative to the speed outlined in the gCode file for a total time of 1 hour and 50 minutes. After printing, the final part was washed in 2:1 tri(propylene glycol) methyl ether/5% acetic acid (3 x 60 mL) to remove unreacted resin and allowed to dry at room temperature in an ambient atmosphere in the dark.

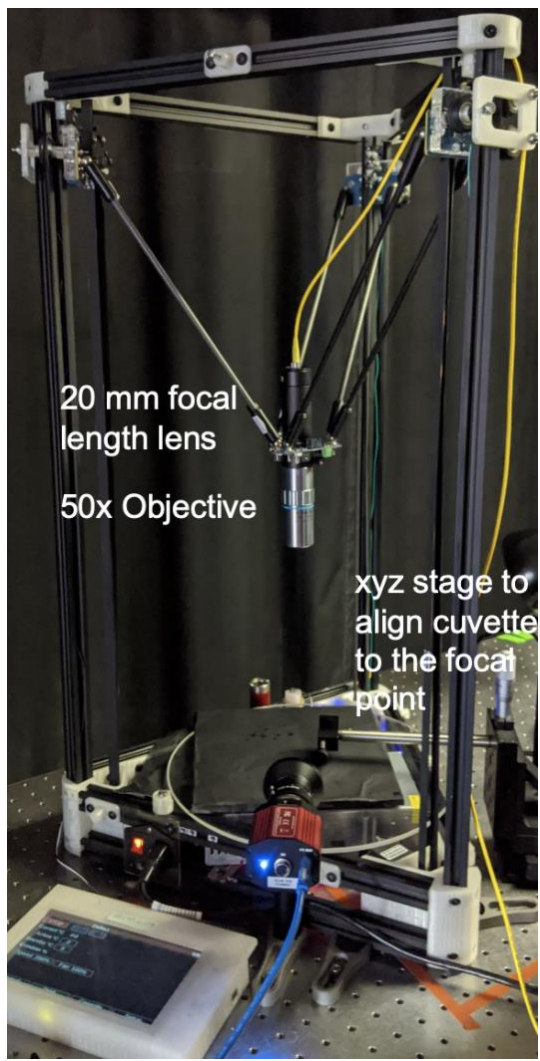

**Figure S1:** Photograph of the fused deposition modeling (FDM) printing setup moves a laser spot in three dimensions. The original instructions for the FDM printer are found at https://www.kosselplus.com/, with our modifications presented above.

**Parallel Excitation Using the DMD Printer:**

TIPS-anthracene Resin Preparation (10 mL): The resin preparation is identical to that of the Br-TIPS-anthracene resin with the following edits.

1. After sparging the resin under nitrogen, the resin was allowed to stir at 45 ºC instead of at room temperature.

2. Gear and Stanford Logo (Fig. 5A, 5B): Instead of pouring the resin into a cuvette, ~0.5 mL of resin was poured into a 35 x 10 mm Falcon brand polystyrene petri dish and well-sealed with parafilm. Pouring a thin layer of the resin into the high surface area petri dish removed bubbles adequately as long as the resin was stirred at 45 ºC after the nitrogen sparge, eliminating the need for the Flacktek SpeedMixer.

3. Lines (Fig. 5C): When printing very small features with light projected through a mask, it was vital to minimize the distance between the top of the mask and the resin. Instead of using a petri dish, a few drops of resin were sealed into a box constructed from 1 in$^2$ glass microscope slide cover slips. To protect the resin from oxygen, the box was constructed the day before, leaving the top open. The edges were sealed with epoxy and cured under UV light. Then, when used for printing, the top was sealed with a small piece of polystyrene plastic, sealed with epoxy. At this stage, the epoxy could not be cured as this would also cure the resin inside of the box. However, this seal was sufficient to print immediately after removing the resin from the glovebox.

The printing is performed on a custom-built setup as shown below. The printing is performed using a custom-built digital micromirror device (DMD) projection, as shown in Fig. S2. The DMD hardware is from Texas Instruments (DLP LightCrafter 6500 Evaluation Module (DLPLCR6500EVM)). The software used to control the DMD is the TI DLP LightCrafter 6500-4.0.1 GUI. Patterns are projected by uploading monochromatic bitmap images using "pattern-on-the-fly" mode. To project the displayed pattern onto the resin, the DMD is illuminated using Luminus Devices CBT-90 Red R5 LED focused using a ThorLabs AL5040M to uniformly illuminate the display area of the DMD. The CBT-90 is driven by a Xantrex (now Ametek/Sorensen) HPD 15-20 current/voltage source and is mounted on a PhlatLight heatsink. The illuminated pattern is reflected from the DMD to a ThorLabs PF20-03-P01 mirror, which reflects the pattern through a ThorLabs LA1433-B. This focused image is then reflected off a ThorLabs PF20-03-P01 mirror and through a ThorLabs AC508-100-A-ML lens which projects the image onto the resin at the lens' focal point.

1. Gear and Stanford Logo (Fig. 5A, 5B): The images of the gear and Stanford logos were projected by the DMD were printed in 8 minutes at a power density of 78 mW cm$^{-2}$. After printing, the final part was washed in 2:1 tri(propylene glycol) methyl ether/5% acetic acid (3 x 60 mL) to remove unreacted resin and allowed to dry at room temperature in an ambient atmosphere in the dark.

2. Lines (Fig. 5C): The lines printed in Fig. 5C were printed in 17 seconds at a power density of 224 mW cm$^{-2}$ using light projected by the DMD through the Negative 1951 United States Air Force Wheel Pattern Resolution Test Targets, 3 x 1 in (Thorlabs).

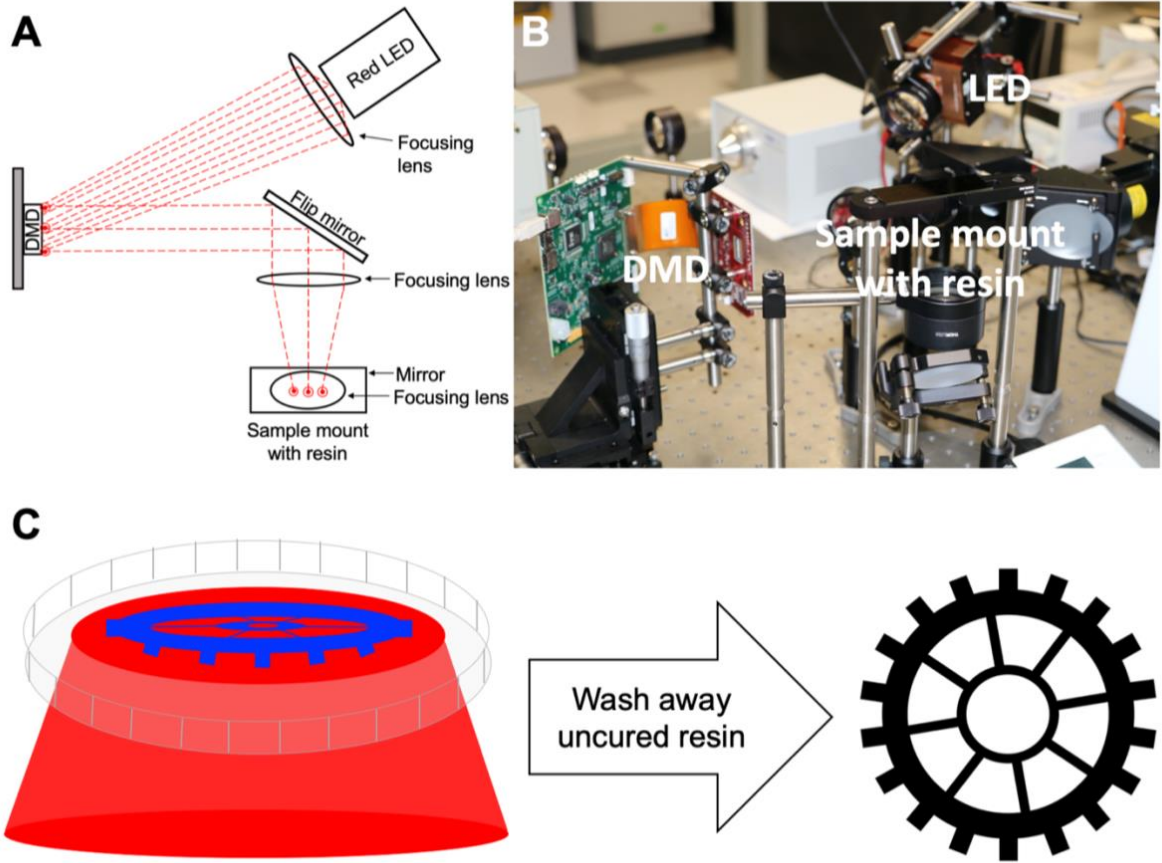

**Figure S2:** Cartoon schematic (a) and photograph (b) of the DMD printing system, which allows for stationary, parallel excitation at one time. (c) Cartoon depiction of the upconversion nanocapsule-facilitated printing process using parallel excitation.

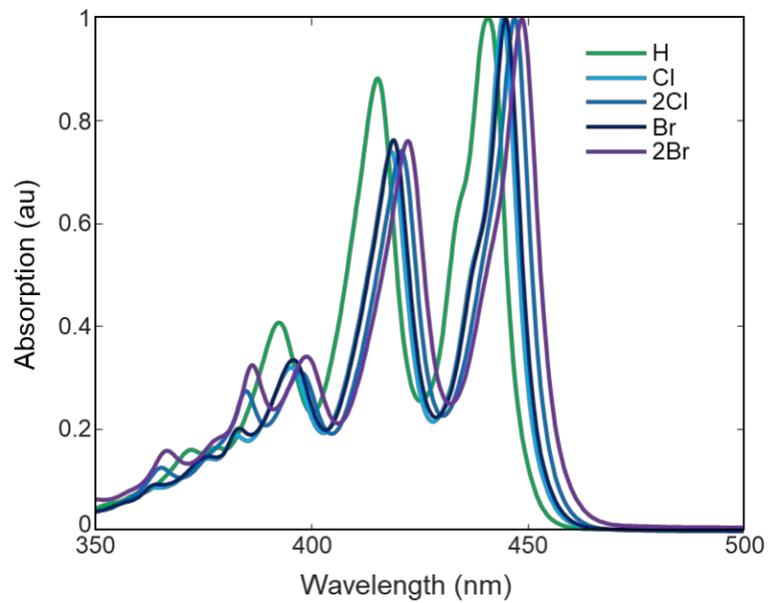

**Figure S3.** Absorption spectra of the annihilators used in this work.

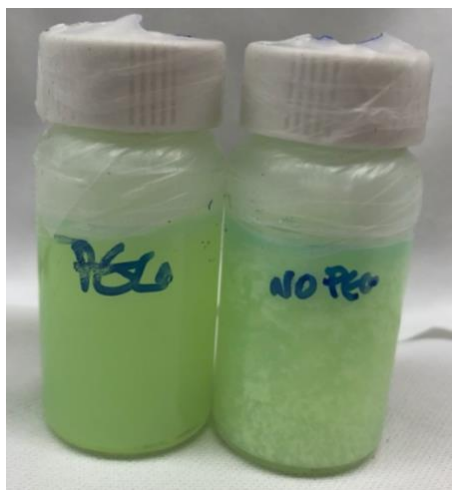

**Figure S4.** Photograph of UCNCs dispersed in water with and without the addition of MPEG-Silane. A precipitate rapidly forms (<1 hour) in the vial without MPEG-silane likely due to nanocapsule aggregation, and this aggregation is irreversible.

| Entry | Effective Diameter (nm) | PDI | Diffusion Coefficient (x $10^{-8}$ cm$^2$ s$^{-1}$) | Data Retained (%) |
|---|---|---|---|---|
| 1 | 76.82 | 0.12 | 6.39 | 99.49 |
| 2 | 76.09 | 0.16 | 6.45 | 98.16 |
| 3 | 77.34 | 0.14 | 6.35 | 99.08 |

**Table S1:** Summary of dynamic light scattering characterization of UCNCs. Each entry represents an average of three measurements from different batches of upconversion nanocapsules synthesized on the scale presented in the Methods Section. This summary highlights the reproducibility of this synthesis on the multi-gram scale.

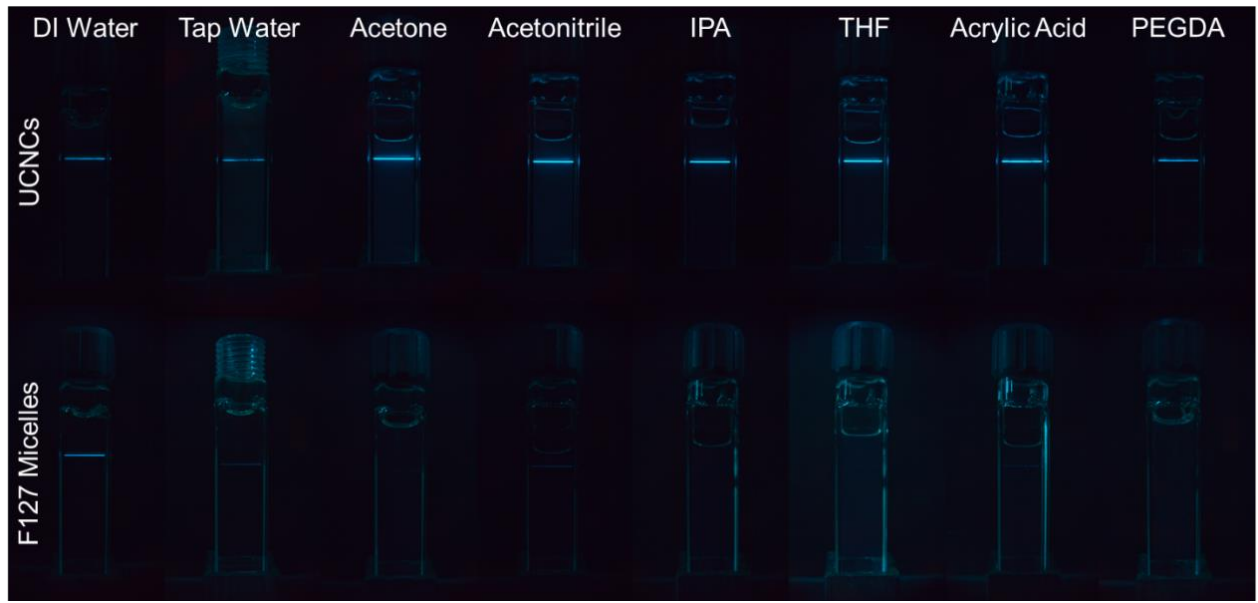

**Figure S5.** UCNCs and F127 micelles dispersed in various solvents. UCNCs and F127 micelles were both synthesized in water and added at 1:30 ratio to the listed solvents. They were then excited at 635 nm and imaged through a 550 nm shortpass filter. The tap water sample was dispersed in water directly from the tap and left uncapped for 20 minutes before taking the image. Acrylic acid and poly(ethylene glycol) diacrylate (PEGDA) were each used to assess capsule durability in acrylate-based monomers for printing resins.

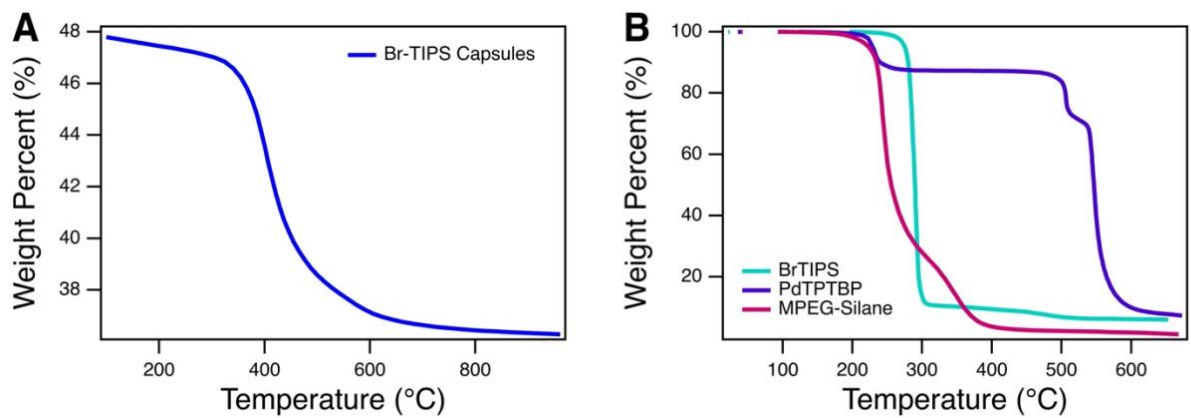

**Figure S6.** Thermogravimetric analysis of (A) capsule paste (B) capsule constituents in nitrogen. In panel (A), the temperature was held at 100 °C until the capsule paste mass remained constant.

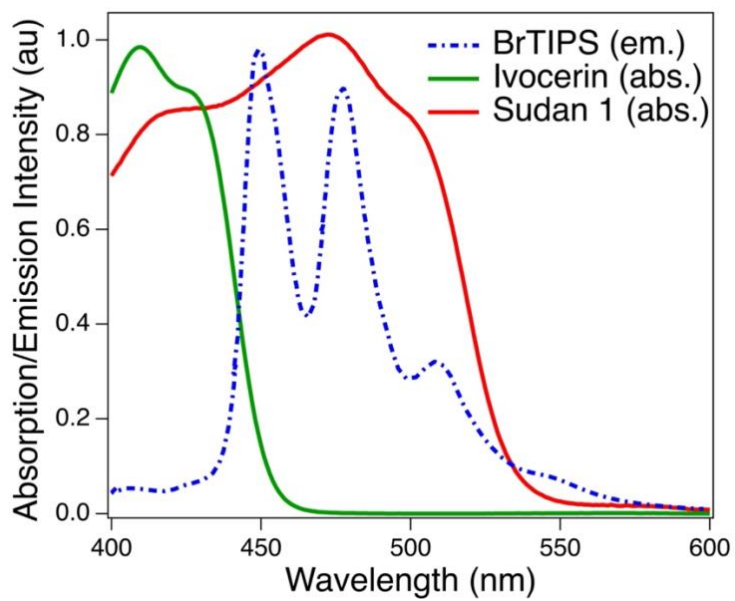

**Figure S7.** Emission-absorption overlap between the upconverted emission (BrTIPS-Anthracene), the photoinitiator (Ivocerin), and the light blocker (Sudan 1).

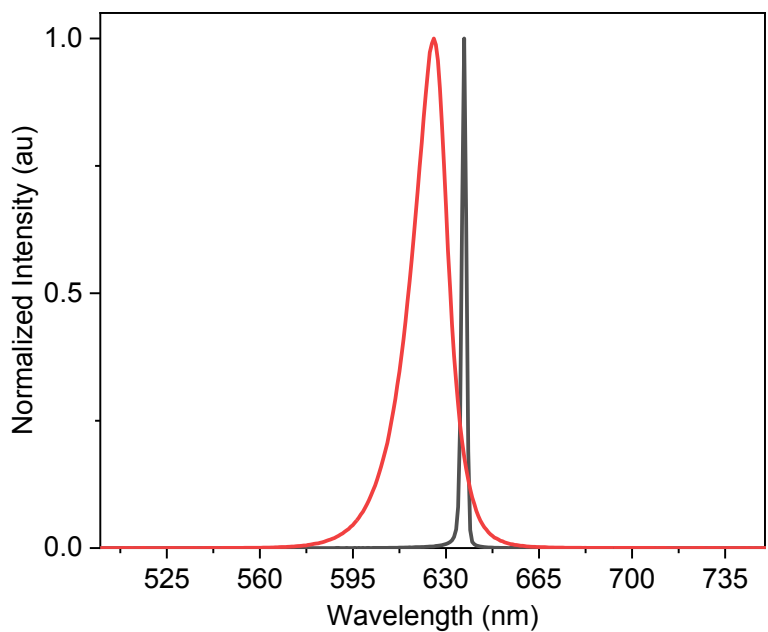

**Figure S8.** Emission spectra of the light sources used to generate prints (black: 637 nm fiber coupled laser for the monovoxel excitation printer (Fig. S1); red: 625 nm LED for the parallel excitation printer (Fig. S2)).

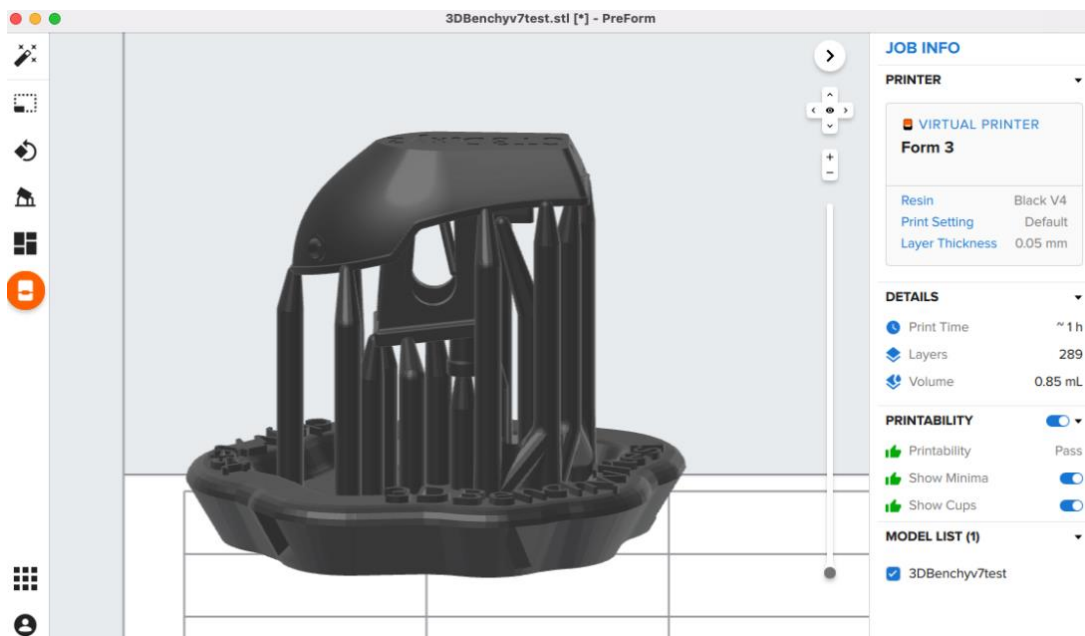

**Figure S9.** Formlabs print of the same Benchy STL. The file was imported into the free software Preform 3.18.0 and simulated for printing on a Form 3B printer at 50 μm layer height. The boat was scaled to match the dimensions of the Benchy printed on our printer. The boat, without support structures, required 200 layers and 0.14 mL of resin. At this point, we used the "one click print" function to generate the printable structure with support structure. This resulted in an object with 289 layers at 50 μm layer height, using 0.85 mL of resin.

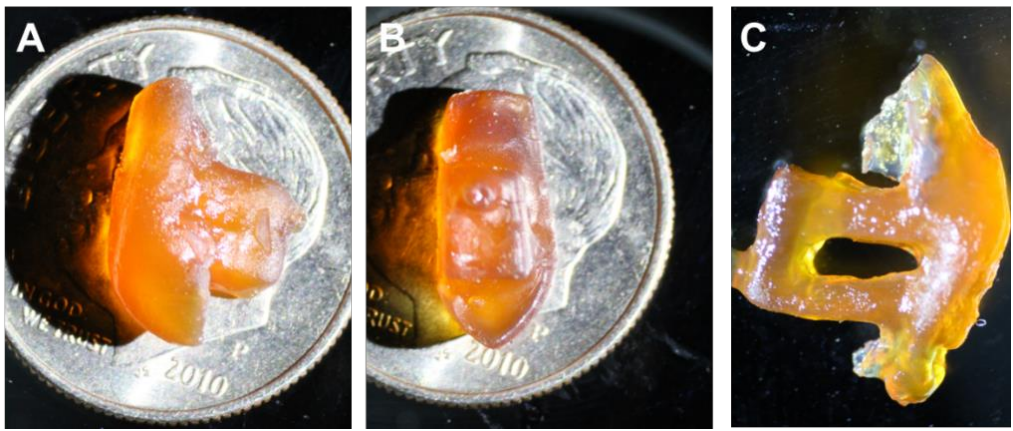

**Figure S10.** An overprinted boat (A,B) gives a lack of discernable features. C) An underprinted boat shows missing features and damage from the wash process. Both issues are remedied by altering the print speed and irradiation power. The boats presented here were printed using a different resin formulation using Bis (5-2,4-cylcopentadien-1-yl)-bis(2,6-difluoro-3-(1H-pyrrol-1-yl)- phenyl) titanium (titanocene) (titanocene, Gelest) as a photoinitiator instead of Ivocerin. This resin was prepared with 1.9 wt% Aerosil 200, 3 wt% titanocene, 13 wt% BrTIPS-Anthracene capsules, 0.03 wt% Sudan 1, and 5 ppm TEMPO. This resin formulation limited printing resolution due to titanocene molecular aggregation, thus the emphasis on the use of the resin presented in the main text and Methods Section.

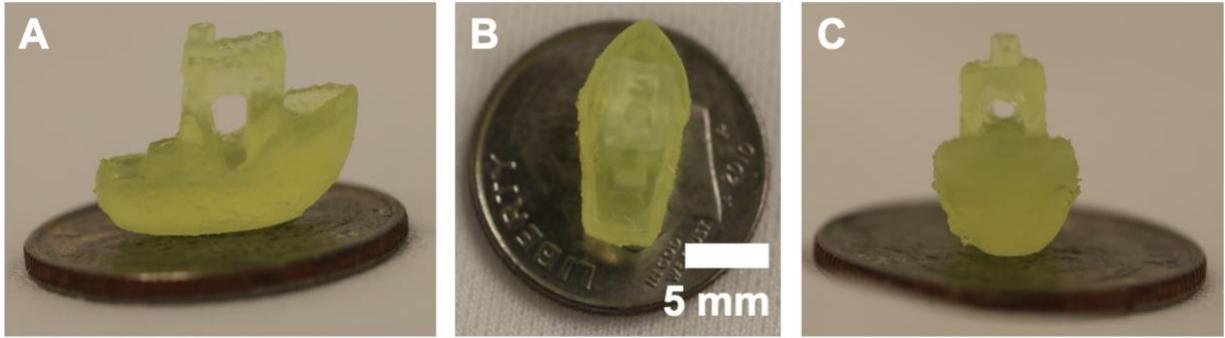

**Figure S11.** The front, side, and top profile images of a Benchy print not presented in the main text (sitting on top of a dime for scale) demonstrate the repeatability of the printing process in our simple, house-made resin presented in the Methods Section.

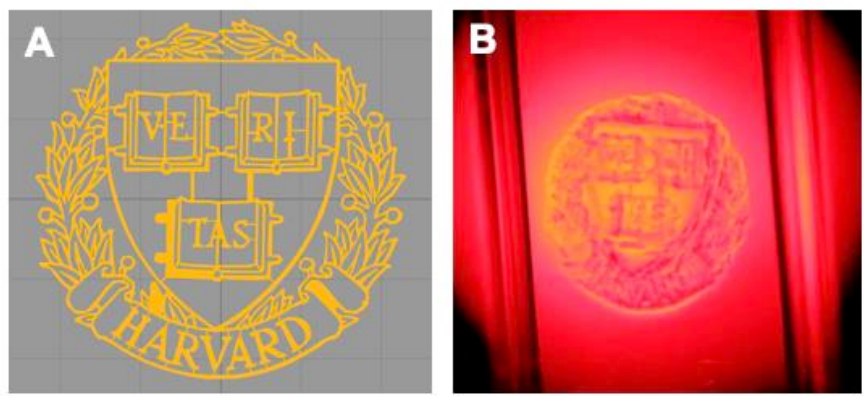

**Figure S12.** The STL file image (a) and a photograph of the Harvard University logo presented in Video S3. The Harvard University Logo presented here was printed using a different resin formulation with Bis (5-2,4-cylcopentadien-1-yl)-bis(2,6-difluoro-3-(1H-pyrrol-1-yl)- phenyl) titanium (titanocene) (titanocene, Gelest) as a photoinitiator instead of Ivocerin. This resin was prepared with 1.9 wt% Aerosil 200, 3 wt% titanocene, 13 wt% BrTIPS-Anthracene capsules, 0.03 wt% Sudan 1, and 5 ppm TEMPO. This resin formulation limited printing resolution due to titanocene molecular aggregation, thus the emphasis on the use of the resin presented in the main text and Methods Section.

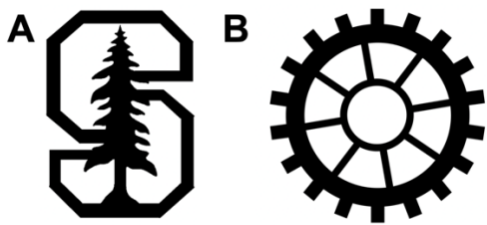

**Figure S13.** The file images projected by the DMD to print the Stanford logo (a) and gear (b) presented in the main text.

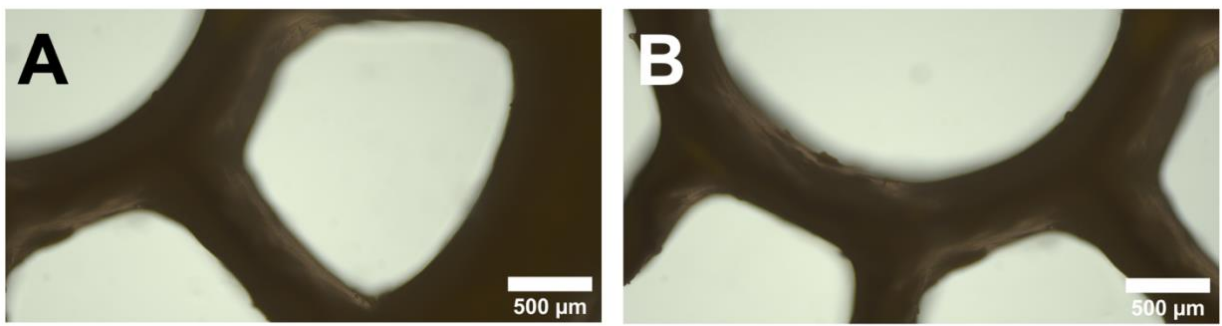

**Figure S14.** Images of a representative gear under the microscope shows that round and straight features are smooth. The images were taken after washing away excess resin and allowing the gear to dry under an ambient atmosphere in the dark.

|  | Column 1 | Column 2 | Column 3 | Column 4 |
|---|---|---|---|---|
| **Actual Width (µm)** | 65.00 | 55.00 | 50.00 | 45.00 |
| **Average Printed (µm)** | 74.35 | 62.68 | 55.30 | 49.51 |
| **Std. Dev (µm)** | 7.99 | 5.10 | 3.95 | 4.84 |

**Table S2:** Quantification of the line widths of the print presented in Fig. 5C. The width of each line was measured in 2-3 places, for a total 18 measurements averaged for lines in columns 1-3. The width of each line was measured in 1-3 places, for a total of 10 measurements averaged for lines in column 4.

**Table S3.** Recent examples of upconversion- or triplet-fusion-facilitated photopolymerization.

| Upconverting Materials | Excitation wavelength | Excitation Power/ Power density | Print Description/Image | Printing Details | Reference |
|---|---|---|---|---|---|
| | | | **Upconversion Using Inorganic Nanoparticles** | | |
| NaYF$_4$:Yb,Tm nanoparticles | 980 nm | 90 W cm$^{-2}$ | 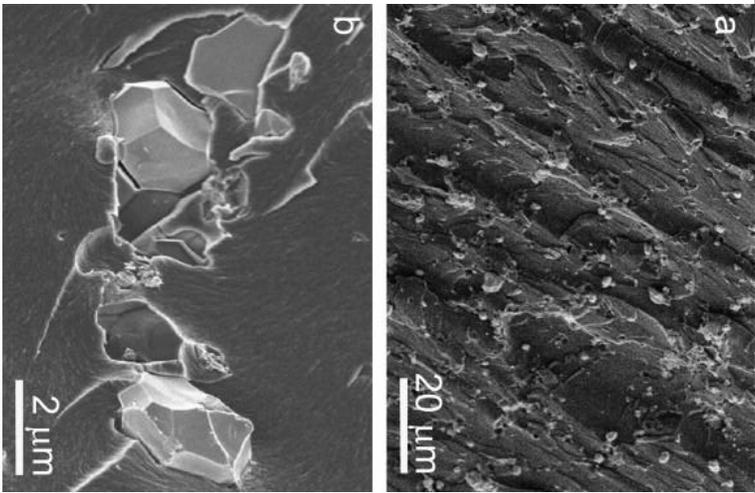 SEM micrographs of polymer/UC phosphor composite: (a) cross-section and (b) magnified view of UC particles incorporated in polymer. Image reprinted from Dental Materials, 28 (3), Alexander Stepuk, Dirk Mohna, Robert N. Grass, Matthias Zehnder, Karl W. Krämer, Fabienne Pellé, Alban Ferrier, Wendelin J. Stark, Use of NIR light and upconversion phosphors in light-curable polymers, 304-311 Copyright 2012, with permission from Elsevier. | Then the samples were molded in cylinders (Ø1 mm) of different thicknesses (1–10 mm). Finally, the premixed and molded specimens were exposed to a NIR laser ($\lambda = 980 \pm 5$ nm, continuous wave, CNI, China) for intervals from 30 s to 5 min. Composite samples of 5 mm thickness were cured two times faster than pure polymer cured by blue light (30 and 60 s, respectively). (max 7-10 mm) | 6 |

| Upconverting Materials | Excitation wavelength | Excitation Power/ Power density | Print Description/Image | Printing Details | Reference |
|---|---|---|---|---|---|
| NaYF$_4$:18%Yb, 0.5%Tm nanoparticles | 980 nm | 9.4 W cm$^{-2}$ | 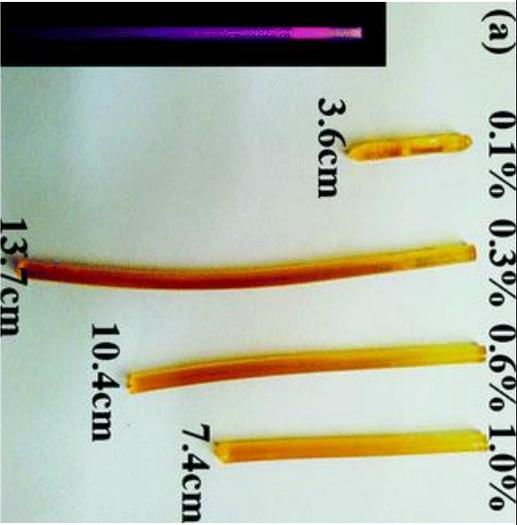 (a) 0.1% 0.3% 0.6% 1.0% 3.6cm 13.7cm 10.4cm 7.4cm  Depths of the cured samples with different concentrations of UCNPs; The inset shows the photopolymerizable sample under excitation by using a 980 nm laser (2 min irradiation). Reproduced from Ref.[7] with permission from The Royal Society of Chemistry. | Then the photocurable samples were injected into a glass tube (17.78 cm long with an outer radius of 5 mm and an inner radius of 4 mm), and then the glass tube was vertically exposed to a fiber coupled laser system (980 nm, Changchun New Industries Optoelectronics Technology). The output power was adjusted to 7.4 W for the activation of the UCNPs. After 2 min irradiation, the cured samples were transferred out of the tube and the uncured parts were washed away by acetone. | 7 |

| Upconverting Materials | Excitation wavelength | Excitation Power/ Power density | Print Description/Image | Printing Details | Reference |
|---|---|---|---|---|---|
| Tm$^{3+}$ doped K$_2$YbF$_5$ nanoparticles | 980 nm | 0.3 W | 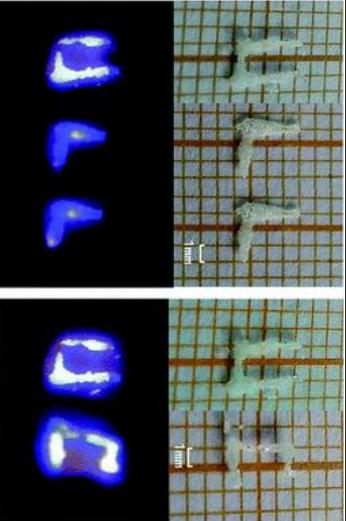3D-printed structures obtained with low-power infrared laser-driven resin curing process as a quasi-instantaneous "laser-writing" procedure. Printed capital letters (upper photographs) correspond to the acronyms ULL, (University of La Laguna) and UC (up-conversion), which show visible luminescence under infrared light radiation (lower photographs) Reproduced from Ref. 8 with permission from The Royal Society of Chemistry. | For this purpose a suspension of micrometre-sized up-converting 0.2 at% Tm$^{3+}$-doped K$_2$YbF$_5$ and a liquid PEGDA resin containing 1.0% in weight of UV-sensitive Irgacure-819® has been prepared in a proportion of 100 g L$^{-1}$. Then, the laser diode describes a previously designed pattern, controlled with an X-Y axis micrometer positioner, even with a continuous moving over the liquid surface (at an estimated speed of 0.3 mm s$^{-1}$) in order to provide local polymerization of the resin. | 8 |

| Upconverting Materials | Excitation wavelength | Excitation Power/ Power density | Print Description/Image | Printing Details | Reference |
|---|---|---|---|---|---|
| NaYF$_4$:Er@Na YF$_4$ Nanoparticles | 1550 nm | See print image | 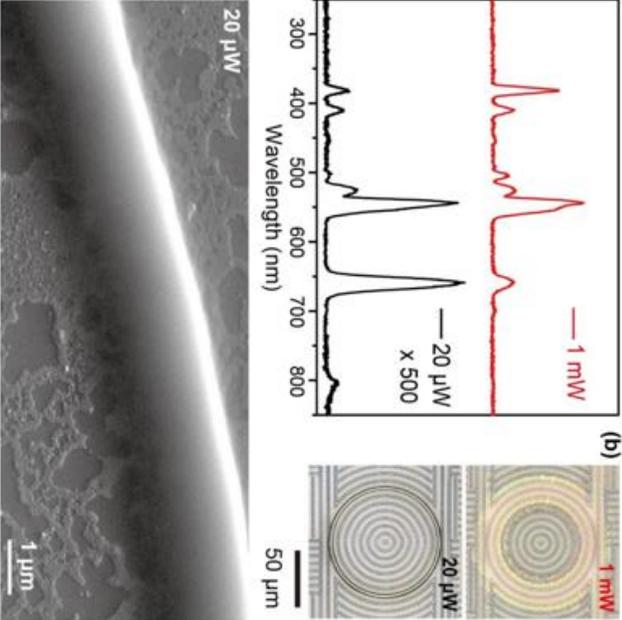(a) Emission spectra recorded under 20 µW and 1 mW excitation, showing the availability of UV emissions at reduced excitation powers. (b) The corresponding SU-8 patterns fabricated under 20 µW and 1 mW excitation, respectively. A relatively low level of excitation is needed for deposition of uniform SU-8 patterns. Excessive UV emission under high power excitation was found to cause uncontrollable photo-curing effect, leading to the formation of irregular SU-8 patterns. (c) 45-degree-tilted scanning electron micrograph shows smooth surface of the SU-8 pattern fabricated under 20 µW excitation. The irregular thin films surrounding the SU-8 pattern are ascribed to residual upconversion nanoparticles. Reproduced from Ref. 9 with permission under CC-BY-4.0. | 20 µL of NaErF$_4$@NaYF$_4$ nanoparticles in ethanol dispersion (0.01 M) was first drop-casted on the microring-resonator. After the ethanol was evaporated, a 2 µm thick SU-8 2002 was spin-coated on the device followed by baking at 90 °C for 10 min. The device was then placed on an alignment stage to align the input/output fiber array with the device. To align the wavelength and polarization of the laser with the microring resonator, we first use a very low input laser power and slowly tune the wavelength as well as adjust the polarization controller until power at the drop port is maximized. Using a very low laser power for finding the resonance wavelength was to avoid too much light exposure to the straight bus waveguide and microring resonator at this stage. Once the resonance and polarization are aligned between the input and the resonator, the laser was then increased to 20 µW to start the curing process. After 10 min the device was removed from the alignment stage and transferred to a 90 °C hotplate for 10 min of post exposure bake procedure. After the device was cooled down to room temperature, then remove the unexposed part of SU-8 by immersing the device into the SU-8 developer for 1 min and finally rinsed in water. | 9 |

| Upconverting Materials | Excitation wavelength | Excitation Power/Power density | Print Description/Image | Printing Details | Reference |
|---|---|---|---|---|---|
| NaYF$_4$:Yb$^{3+}$,Tm$^{3+}$/NaYF$_4$ nanoparticles | 975 nm | 7 W cm$^{-2}$ | 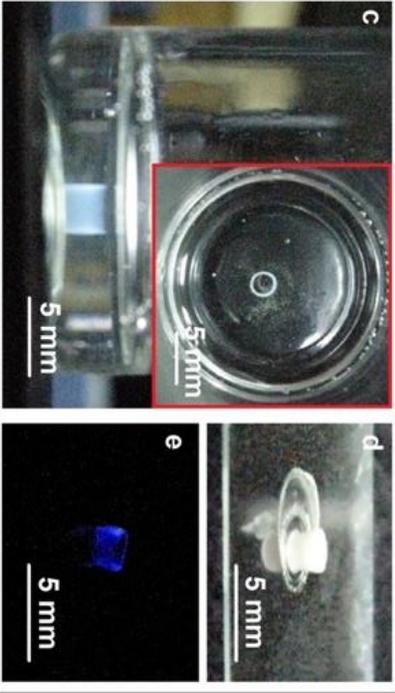 (c) Top-view image of a structure produced by NIR triggered 3D photopolymerization, (d) a 3D structure obtained via NIR-triggered photopolymerization after developing and (e) its anti-Stokes luminescence under 975 nm excitation. Reproduced from Ref.[10] with permission under CC-BY-4.0 | Commercial CW laser diode at 975 nm (ATC-SD, Russia) was focused by an objective in PCC volume, containing UCNPs. The power density of the laser was set at 15 W cm$^{-2}$. As a proof of feasibility of the proposed approach for creation of the 3D printed device macroarchitecture, the hollow tube formation with the diameter of 1.5 mm and the height of 2 mm was obtained by layer-by-layer drawing directly in the resin volume. The PCC was scanned by the laser via 2-axis galvano-scanner Miniscan-07 (Raylase, Germany). The displacement of voxel along the z-axis was carried out by a micrometric translation stage (Thorlabs). After finishing of the drawing process, the tube was removed and developed in 2-propanol. | 10 |

| Upconverting Materials | Excitation wavelength | Excitation Power/ Power density | Print Description/Image | Printing Details | Reference |
|---|---|---|---|---|---|
| β-phase NaYF₄:TmYb @NaYF₄ nanoparticles (core = NaYF₄: 0.5 mol% Tm³⁺, 30 mol% Yb³⁺; shell = NaYF₄) | 974 nm | 0.64 W cm⁻² | 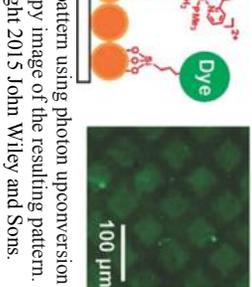 a) Procedure for the fabrication of a fluorescent pattern using photon upconversion lithography. b) Confocal laser scanning microscopy image of the resulting pattern. Reproduced with permission from ref. [11]. Copyright 2015 John Wiley and Sons. | For the patterning of proteins, a photomask was positioned between the NIR light and the proteins. In the exposed areas, UCNPs convert NIR light into blue light that induces cleavage of the Ru complexes and the local release of proteins. The Ru-UCNP substrate was irradiated with 974-nm light (0.64 W/cm²) for 40 minutes. | 11 |
| NaYF₄: Yb³⁺, Tm³⁺ nanoparticles | 980 nm | 3–6 W | 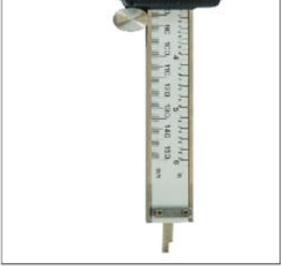 Cured samples containing 20% UCP. Image reprinted from Progress in Organic Coatings, 104, Masoume Kaviani Darani, Saeed Bastani, Mehdi Ghahari, Pooneh Kardar, Ezeddin Mohajerani, NIR induced photopolymerization of acrylate-based composite containing upconversion particles as an internal miniaturized UV sources, 97-103 Copyright 2017, with permission from Elsevier. | The UV-curable composition were placed in a cavity inside a glass panel. The sample containing acrylate monomer, photo-initiator (PI) and UCPs was exposed to NIR radiation cum disparate intensity. | 12 |

| Upconverting Materials | Excitation wavelength | Excitation Power/ Power density | Print Description/Image | Printing Details | Reference |
|---|---|---|---|---|---|
| βNaYF$_4$:18%Yb,0.5%Tm nanoparticles | 980 nm | 19.3 W | 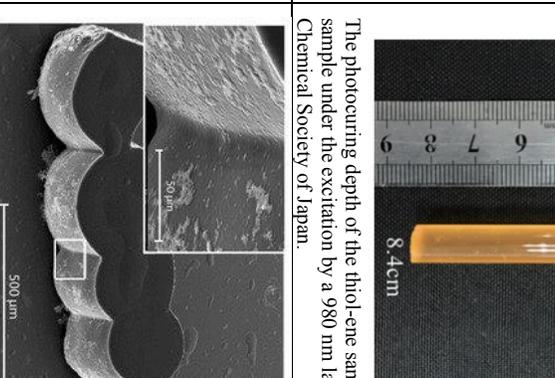The photocuring depth of the thiol-ene sample. The inset shows the photopolymerizable sample under the excitation by a 980 nm laser. Reproduced with permission from the Chemical Society of Japan. | The formulation contained 0.3 wt % of UCNPs, 0.3 wt % of photoinitiator Irgacure 784, and equal molar ratio of thiol-ene. The photosensitive resin was injected into a glass tube, which was then vertically exposed to a 980 nm laser. The output power of the laser and the exposure time were 19.3 W and 40 s, respectively. The cured samples were transferred out of the tube and the uncured parts were washed away with acetone. | 13 |
| β-NaYF$_4$:Yb$^{3+}$, Tm$^{3+}$ nanoparticles | 975 nm | ~ 100 W/cm$^2$ | SEM image of 3D polymer microstructures obtained by NIR light-activated photopolymerization: (**a**) UCNPs concentration ~ 20 mg mL$^{-1}$ and (**b**) UCNPs concentration ~ 2 mg mL$^{-1}$. Reproduced from Ref. [14] with permission under [CC-BY-4.0](#) | A mixture of oligocarbonate methacrylate (OCM-2), photoinitiator Irgacure 369, and nanoparticles was used as the photocurable composition for the UCNP-assisted NIR polymerization in bulk. First, 10 mg of Irgacure 369 was mixed with 1 g OCM-2, then, 100 μL UCNP dispersed in hexane (c = 0.2 g/mL) was added and the mixture was thoroughly shaken and sonicated until the hexane evaporated. The photopolymerization occurred in a laser beam (975 nm) focused into a specific volume of glass vial containing the photocurable composition at a laser power of ~ 100 W/cm$^2$ for 20 s. | 14 |

## Triplet Fusion Upconversion

| Upconverting Materials | Excitation wavelength | Excitation Power/ Power density | Print Description/Image | Printing Details | Reference |
|---|---|---|---|---|---|
| TTA-upconversion system: [IBDP-IBDP] = 1.0 × $10^{-5}$ M, [Perylene] = 1.0 × $10^{-4}$ M in nanomicelles prepared according to [15]. | 635 nm | 4.6 mW | 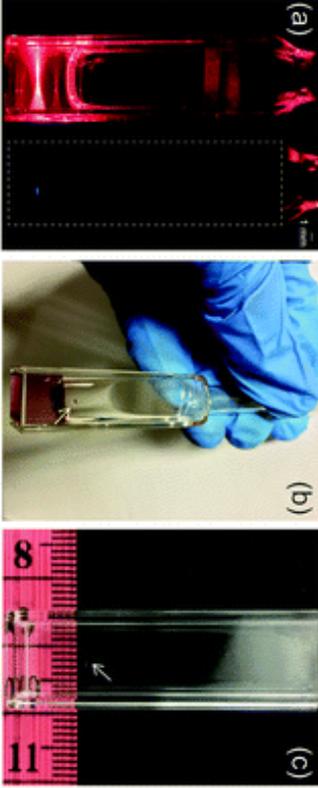 Spatially confined photopolymerization driven by TTA upconversion. (a) Photographs of TTA-upconversion in the nanomicelle, $\lambda_{ex}$ = 635 nm, laser power 4.6 mW. The right part of the panel is the photograph taken with a 400–520 nm bandpass filter. The localized excitation is indicated by the blue upconversion emission spot. (b) and (c) Application of the TTA upconversion nanomicelle-based localized photoexcitation in the spatially confined photo-polymerization (the polymer particle is indicated by the arrows in (b) and (c). Note the polymerization only occurred at the focusing point of the laser beam). Reproduced from Ref. [16] with permission from The Royal Society of Chemistry. | Photopolymerization system: monomer: 2-hydroxylethyl acrylate (HEA); photoinitiator: Irgacure 784, irradiated with 635 nm cw laser (2 mW, power density at a focal point is 0.26 W $cm^{-2}$) irradiation for 40 s. | 16 |
| Silica encapsulated upconversion nanocapsules containing PdPTBP and TIPS-Anthracene or PdPTBP and Br-TIPS-Anthracene | 625 nm LED  637 nm fiber coupled laser | LED: 78 mW $cm^{-2}$ - 224 mW $cm^{-2}$  Laser: 4.5x10³ - 4.9x10³ W $cm^{-2}$ | 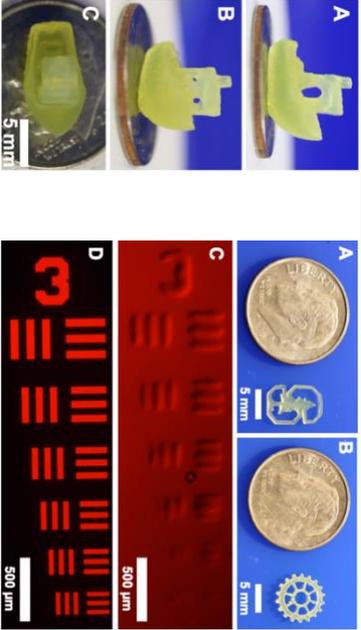 | For each measurement, Irgacure 784 (2 mg) was added into 0.5 mL HEA (heating with dryer to fully solubilized it), then nanomicelle mother solution (0.5 mL) was added. The mixture was filtered out by Syringe-driven Filter (Nylon, 0.22 μm) to obtain the clear and transparent nanomicelle photopolymerization solution for measurement.  See Methods Section. | This work. |

**Supplemental Note 1:**

Approximating the distance from the focal point where printing occurs in the quadratic regime.

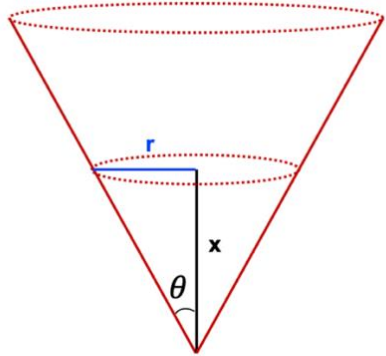

Input light focused by the 0.55 NA objective used for monovoxel-based printing
r : Radius of the voxel cross section at the quadratic threshold
x : Distance from the focal point where printing occurs in the quadratic regime

1) Determine the area ($A$) of the voxel cross section at the quadratic threshold for BrTIPS-Anthracene. The 4 mW is the measured power ($P$) at the focal point, as reported in the monovoxel printing methods section, and the 121,100 mW cm$^{-2}$ ($I_{th}$) is the quadratic threshold presented in Fig. 2.

$$A = \frac{P}{I_{th}} = 4 \; mW * \frac{1 \; cm^2}{121,100 \; mW} * \frac{(10^4 \; \mu m)^2}{(1 \; cm)^2} = 3,300 \; \mu m^2$$

2) Determine the radius of the voxel cross section at the quadratic threshold (Fig. 2) for BrTIPS-Anthracene.

$$A = \pi r^2$$

$$3,300 \; \mu m^2 = \pi r^2$$

$$r = 32.4 \; \mu m$$

3) Determine the angle ($\theta$) using a 0.55 NA objective and approximating the index of refraction ($n$) of 1.5 for the resin.

$$NA = n \sin(\theta)$$

$$0.55 = 1.5 \sin(\theta)$$

$$\theta = 21.5°$$

4) Determine the distance from the focal point that printing occurs in the quadratic regime based on the radius of the voxel cross section determined in (2).

$$x = (32.4 \ \mu m) \cot(21.5°)$$

$$x = 82 \ \mu m$$

**List of Supplemental Files:**

**Supplemental File 1.** The STL file uploaded to Simplify 3D to generate the Benchy print.

**Supplemental File 2.** The gcode file used to generate the Benchy print presented in this manuscript. The STL file was imported into the software Simplify3D, the dimensions of Benchy were scaled, and exported as a gcode file to control the monovoxel excitation printer presented in Fig. S1.

**Video S1.** A relative viscosity comparison of resin with Aerosil 200 (R, right/bottom cuvette) with PETA (P, left/top cuvette).

**Video S2.** Time lapse video of the monovoxel excitation-based printing of Benchy using the Br-TIPS-anthracene resin. The photograph at the end of the video shows representative contents of a cuvette containing the Benchy print along with uncured resin to be later washed off of the print. The actual time to generate the print is 1 hour 50 minutes.

**Video S3**. Video of a front and profile view of the Harvard logo printed inside a 1 cm pathlength polystyrene cuvette with the FDM printer.

**Video S4.** Time lapse video of the DMD-based printing of a gear with the TIPS-anthracene resin, with a photograph at the end showing the contents of the petri dish after a print. The actual time to generate the print is 8 minutes.